FRONT MATTER

Title : Suppressed terahertz dynamics of water confined in nanometer gaps

Authors

Hyosim Yang[1,†], Gangseon Ji[1,†], Min Choi[2], Seondo Park[3], Hyeonjun An[1], Hyoung-Taek Lee[1], Joonwoo Jeong[1], Yun Daniel Park[3], Kyungwan Kim[4], Noejung Park[1], Jeeyoon Jeong[5,*], Dai-Sik Kim[1,3,*], and Hyeong-Ryeol Park[1,*]

Affiliations

1. Department of Physics, Ulsan National Institute of Science and Technology, Ulsan 44919, Republic of Korea
2. Department of Chemistry, Ulsan National Institute of Science and Technology, Ulsan 44919, Republic of Korea
3. Department of Physics and Astronomy, Seoul National University, Seoul 08826, Republic of Korea
4. Department of Physics, Chungbuk National University, Chungbuk 28644, Republic of Korea
5. Department of Physics and Institute for Quantum Convergence Technology, Kangwon National University, Gangwon 24341, Republic of Korea

[†]These authors contributed equally to this work: Hyosim Yang and Gangseon Ji.

**Contact details of the corresponding author**[*]
    **Prof. Hyeong-Ryeol Park**

    Department of Physics, Ulsan National Institute of Science and Technology, Ulsan, 44919, Republic of Korea
    [*]E-mail: nano@unist.ac.kr
    **Prof. Dai-Sik Kim**

    Department of Physics, Ulsan National Institute of Science and Technology, Ulsan, 44919, Republic of Korea / Department of Physics and Astronomy, Seoul National University, Seoul, 08826, Republic of Korea
    [*]E-mail: daisikkim@unist.ac.kr
    **Prof. Jeeyoon Jeong**

    Department of Physics and Institute for Quantum Convergence Technology, Kangwon National University, Gangwon 24341, Republic of Korea
    [*]E-mail: peterjjy@kangwon.ac.kr

**E-mail** : Hyosim Yang (hyyang3070@unist.ac.kr), Gangseon Ji (gangseon@unist.ac.kr), Min Choi (minc.3232@gmail.com), Seondo Park (sheldon17@snu.ac.kr), Hyeonjun An (rnwk23@unist.ac.kr), Hyoung-Taek Lee (htaek@unist.ac.kr), Joonwoo Jeong (jjeong@unist.ac.kr), Yun Daniel Park (parkyd@snu.ac.kr), Kyungwan Kim (kyungwan@cbnu.ac.kr) Noejung Park (noejung@unist.ac.kr), Jeeyoon Jeong (peterjjy@kangwon.ac.kr), Dai-Sik Kim (daisikkim@unist.ac.kr), and Hyeong-Ryeol Park (nano@unist.ac.kr)

**Abstract**
Nanoconfined waters have been extensively studied within various systems, demonstrating low permittivity under static conditions; however, their dynamics have been largely unexplored due to the lack of a robust platform, particularly in the terahertz (THz) regime where hydrogen bond dynamics occur. We report the THz complex refractive index of nanoconfined water within metal gaps ranging in width from 2 to 20 nanometers, spanning mostly interfacial waters all the way to quasi-bulk waters. These loop nanogaps, encasing water molecules, sharply enhance light-matter interactions, enabling precise measurements of refractive index, both real and imaginary parts, of nanometer-thick layers of water. Under extreme confinement, the suppressed dynamics of the long-range correlation of hydrogen bond networks corresponding to the THz frequency regime result in a significant reduction in the terahertz permittivity of even 'non-interfacial' water. This platform provides valuable insights into the long-range collective dynamics of water molecules which is crucial to understanding water-mediated processes such as protein folding, lipid rafts, and molecular recognition.

**Teaser**
The suppressed vibrational modes in nanoconfined water lead to an apparent solid-like behavior at terahertz frequencies as observed experimentally.

**MAIN TEXT**

**Introduction**

Water molecules behave differently under nanoconfinement, which significantly affects various processes in our everyday life such as solvation of ions and biomolecules (*1-8*), molecular transport through nanopores (*9, 10*), formation of electric double layer on electrodes (*11, 12*), and chemical reactions (*13, 14*). Accordingly, understanding the properties of nanoconfined water is of paramount importance in all fields of natural science and has attracted many theoretical and experimental studies (*15-33*). Thus far, most studies have agreed on the rule of thumb that the nanoconfinement effect comes from the unique response of interfacial water, which forms an ordered state parallel to the interface up to 1 nm thickness (~ 4 layers of water molecules), leading to a suppressed reorientation of molecular dipole moment under external field bias and consequently a decreased permittivity (*34-37*).

Despite decades of efforts to experimentally determine permittivity of interfacial water, mostly in naturally occurring nanoporous materials (*9, 23, 38*), it was only recently that permittivity of nano-water has been measured in a systematic way (*39*). In this seminal work, a van der Waals assembly of two-dimensional (2-D) materials was used to precisely control the thickness of water, and the permittivity is measured with an electrostatic force microscopy (EFM), which provides sub-piconewton resolution to detect molecules with ultralow polarizabilities.

Static permittivity of nanoconfined water has been studied extensively in the last few years; however similar progresses in its high frequency counterpart have been lacking. Most optical measurements on nanoconfined water have been performed in systems whereby a sufficient surface area can be provided, such as with beryl crystals (*40, 41*), silica aerosols (*9, 42*) and metallic waveguides (*43-45*), yet without a reliable way to control the thickness of water below 10 nm regime. Thus far, the best way to control thickness of water at nanometer scale is leak-controlled deposition of water vapor under ultra-high vacuum and cryogenic temperatures (*46, 47*). However, to be relevant to biological phenomena, where optical characterization and manipulation of biomolecules are becoming increasingly important, it would be ideal to have a platform where optical properties of nano-water can be systematically studied under ambient environment (*48*).

Optical nano-antennas are ideal for such purpose, as it can provide both nano-confinement and an increased sensitivity at the same time. Especially, loop nanogap antennas support strong dipolar resonance and transmit light only through the gap without background transmission (*48-51*), allowing us to analyze light-matter interaction exclusively in the gap (*52-55*). Recently, it has also been demonstrated that liquid water can be filled into 10 nm-wide loop nanogaps by replacing a spacing dielectric layer with water molecules via etch-and-dilute process (*44*). Since the gap width is determined by thickness of the spacing dielectric layer (*48, 56*), which can be nanometrically controlled in the deposition step, the loop nanogaps are ideal for systematic investigation of nanoconfined water at different thicknesses. Additionally, the loop nanogap antenna, which can be fabricated on a wafer scale using atomic layer lithography, enables optical measurements of the vertically oriented nanometer-thick water layers in the long wavelength regime to investigate long-range hydrogen bond networks, which couldn't be reached previously.

In this study, we optically measured water-filled rectangular loop nanogaps with gap widths of 2 to 20 nm to determine refractive indices of nanoconfined water at low terahertz (THz) frequencies between 0.1 and 1.5 THz, from which one can access long-range collective dynamics responsible for solvation dynamics (*57-61*) (Fig. 1A). By way of the high sensitivity of loop nanogaps, together with the small spectral dispersion of water at THz frequencies, water-induced changes in the resonant frequencies and amplitudes of the loop nanogaps could be unambiguously converted into complex refractive indices of the gap-filling nanoconfined water (*44*). The retrieved complex refractive indices of nanoconfined water were anomalously low compared to their bulk counterparts, being even lower at smaller gap widths. At the gap width of 2 nm, the real index was only 60 % of the bulk value, and the imaginary index nearly approached zero. From complementary analyses, comprising ab-initio calculation, coupled mode calculation, and effective medium theory, the suppressed Debye relaxation dynamics of quasi-bulk water at the center (Fig. 1C), which are not quite bulk-like despite being 10 nm away from the interface, explain the overall anomalously low complex refractive indices of the nanoconfined water at THz frequencies.

## Results
### Preparation of nanoconfined water in metal gaps with various widths

Water-filled nanogap loops are realized in two steps; first, metal-insulator-metal gaps are fabricated using atomic layer lithography (*48*), and then the gap-filling insulator is substituted with liquid water using a wet-etch-and-dilution process (*44*). First, a 200-nm-thick gold layer with rectangular holes is patterned on a 500 μm-thick high-resistivity silicon wafer using conventional photolithography (Fig. 2A). In Fig. 2B, aluminum oxide is deposited conformally over the entire metallic pattern using atomic layer deposition (ALD) with an angstrom-scale precision. Aluminum oxide thickness varies from 2 to 20 nanometers. After the ALD process, a 200 nm-thick gold layer is secondly deposited to fill the rectangular hole, and the excess gold layer on top of the first metal pattern is exfoliated using an adhesive tape. Thus, vertically aligned nanogaps between the metals are formed, whose width is determined by the thickness of aluminum oxide deposited using ALD. The nanogap fabrication is completed by polishing the surface using Ar ion milling at an oblique angle (Fig. 2C). Next, to wet-etch the aluminum oxide on top of and inside the gap, nanogap samples are immersed in 1 M potassium hydroxide (KOH) solution for 1 to 10 minutes, depending on the gap width (Fig. 2D). Then, the samples are transferred to a deionized (DI) water bath (R = 18.2 MΩ·cm) for dilution process. The wet-etch residues were completely replaced with water molecules after more than an hour (Fig. 2E). In order to prevent gap filling water from vaporizing, a sapphire cover is attached to the sample with a 100 μm-thick double-sided Kapton tape as a spacer (Fig. 2F). The reservoir is further sealed on the sides with GE varnish during temperature-dependent measurements in a cryostat, preventing the gap-water from leaking. It should be noted that the gap is always filled with aluminum oxide, KOH or water, so that liquid water can be guaranteed to fill the gap after the whole process is completed. An empty gap is prepared as a reference by vaporizing the gap-filling water by critical point drying (CPD) (*62*). Detailed fabrication processes are shown in Methods. Cross-sectional scanning transmission electron microscopy (TEM) images and energy dispersive spectroscopy (EDS) maps of $Al_2O_3$-filled and empty nanogaps (Fig. 2G and 2H) suggest that the wet-etch process successfully removes aluminum oxide without damaging the gap, further confirming that our nanogaps are completely filled with water.

### THz transmissions through empty and water-filled nanogaps

The nanogaps are characterized by terahertz time-domain spectroscopy (THz-TDS), as detailed in Methods. In Fig. 3A, the time-domain transmission signals of empty and water-filled nanogaps as well as their respective reference signals are presented. It is evident from the time traces of narrower gaps that oscillation periods are longer than those of their 10 nm-gap counterparts, implying that resonances exist at lower frequencies due to stronger gap plasmon coupling at narrower gaps. Upon filling the gap with water, we observe a significant suppression of the tailing oscillation, which results from absorption introduced by the water filling the gap. In Fig. 3B, the corresponding frequency-domain transmission spectra are shown after Fourier-transforming the time traces in Fig. 3A. The introduction of water into the nanogap causes a redshift in all the resonant peaks of the samples and a decrease in peak amplitude. Noticeably, the decrease in amplitude is less pronounced in narrower gaps, which is counterintuitive since narrower gaps tend to exhibit larger spectral responses to changes in the dielectric environment near the gap. This is also evident from the blue dashed lines in Fig. 3B, which are analytically calculated spectra using the coupled-mode method with the refractive indices of bulk water (please refer to the details of coupled-mode method and THz complex refractive indices of bulk water in Supplementary Information Fig. S1 and S2, respectively). These experimental results suggest that the gap-filling water can have a lower dielectric permittivity in THz region when the gap is narrower, similar to the earlier results in static and infrared (IR) regimes (*39, 45*).

**Estimation of THz complex refractive index of nanoconfined water**

In order to quantitatively estimate refractive indices of the gap-filling water in THz region, we calculated transmission spectra of the nanogap sample (see Supplementary Materials Fig. S3) and created two-dimensional (2D) maps of the relative peak amplitudes and relative resonance frequencies as a function of complex refractive indices ($\tilde{n} = n + i\kappa$) of the gap-filling medium using the coupled-mode method, as shown in Fig. 4A and 4B. It should be noted that the relative peak amplitude and relative resonance frequency are normalized with respect to the values obtained from an empty gap. Since the resonance frequency is predominantly affected by the real part of the refractive index $n$, and the transmitted amplitude by the extinction coefficient $\kappa$ (Fig. S3), one can unanimously determine $(n, \kappa)$ of the gap-filling water by analyzing the relative changes in peak amplitude and resonant frequency of the corresponding transmission spectrum. For example, a water-filled gap with a width ($w$) of 5 nm and a loop length ($L = 2(L_x + L_y)$) of 200 μm, as shown in Fig. 4A and 4B, exhibits a relative peak amplitude and resonance frequency of 63% and 73%, respectively, compared to an empty gap, representing a complex refractive index of $\tilde{n} = 1.5 + i0.082$ for the nanoconfined water at the resonance of the loop nanogap. In spite of the same loop length, the resonance peak in the spectrum varies when passing through a narrow gap of tens of nanometers or less, because of the gap plasmon effect (*50*). Interestingly, this has been also experimentally validated using a smaller loop nanogap with a loop length of $L = 100$ μm, resulting in different resonance frequencies (see Supplementary Materials Fig. S4), while maintaining the same gap sizes at 5 and 20 nm. Figures 4C and 4D summarize the results of such analyses at various gap widths, where $(n_\perp/n_{bulk}, \kappa_\perp/\kappa_{bulk})$ are relative values to the refractive indices of bulk water at each resonance frequency. Even though water confined in narrower gaps shows smaller refractive indices, the loop length does not significantly affect the trend, indicating that the observed phenomena are the result of confinement and are minimally affected by dispersion.

There are a number of reasons why water within the nanogap has a lower refractive index (*18, 39, 45, 58, 63*). One of these is the low refractive index of the interfacial water formed at the surface of the nanogap sidewall. An interfacial water layer represents an ordered layer of water molecules, whose rotation is suppressed at the solid-liquid interface. The thickness of this interfacial water layer can be influenced by the hydrophilic or hydrophobic nature of the sidewall of the metal nanogap. Due to its hydrophilic nature of the gold surface of the nanogap (see Supplementary Materials Fig. S5), an estimated thickness of the interfacial water layer is about 0.75 nm, as calculated by a hybrid-type approach with molecular dynamics (MD) and density functional theory (DFT) calculations (see Supplementary Materials Fig. S6). Therefore, the spectral response of a 2 nm gap will be dominated by the refractive indices of the interfacial water whose thickness is 0.75 nm × 2 = 1.5 nm; $\tilde{n}_\perp$ estimated from the 2 nm gap sample sets an upper limit for the refractive index of the interfacial water, i.e., $\tilde{n}_{upper} = 1.41 + i0$. If we follow the convention to use refractive index of water at visible frequencies (*57*) as the lower bound $n_{lower} \simeq \sqrt{\varepsilon_\infty} = 1.33$, we continue our discussion assuming $\tilde{n}_i = 1.37 + i0$ as a representative value for the refractive index of interfacial water.

With the assumption on the refractive index of the interfacial water, we can evaluate refractive indices of the residual water in all gap widths based on the effective medium theory (*64*) (Fig. 5A), which was also used to evaluate permittivity of nanoconfined water in the static regime (*39*). Following this approach, we first assumed that the gap-filling water comprises of (1) 0.75 nm-thick interfacial water layers on both sides of the gap, and (2) the residual water with bulk properties. Then, we calculated the effective dielectric permittivity of the nanoconfined water as $\tilde{\varepsilon}_\perp = w/[2x/\varepsilon_i + (w-2x)/\tilde{\varepsilon}_{res})]$ where $x$ is the thickness of the interfacial water, $\varepsilon_i$ is the dielectric permittivity of the interfacial water, and $\tilde{\varepsilon}_{res}$ is that of the residual water sandwiched between the interfacial water (*64*). We utilized the relationship $\tilde{\varepsilon} = \tilde{n}^2$ for the effective medium theory calculation, such that $\varepsilon_i = \tilde{n}_i^2 = 1.9$. Results of the effective medium analyses are summarized in Fig. 5B and 5C, which show significant differences between the refractive indices in Fig. 4C and 4D and the effective refractive indices derived by applying effective medium theory (see Supplementary Materials Fig. S7). Furthermore, these results allow us to determine $\tilde{n}_{res}$ in each sample, and we observe large deviations of $\tilde{n}_{res}$ from the refractive index of bulk water (see Supplementary Information Tables S1 and S2), which implies that there are additional mechanisms that affect the dynamics of residual water at THz frequencies, which differs from static conditions (*39*), as shown in Fig. 5D to 5F.

In order to better understand the underlying physics of residual water dynamics at THz frequencies, we refer to some previous reports which suggest that molecular dynamics at THz frequencies follows a much larger characteristic length scale. In Popov et al. and Elton et al. (*59, 60*), the authors report an existence of a fast Debye process (~1 ps) operating between 0.1 and 1 THz, in addition to the slow Debye process at 18 GHz (*57*). The fast Debye process is primarily a result of the vibration of the hydrogen bond network, or more specifically, the rotation of weakly bound water molecules. Due to the delocalized nature of the hydrogen bond network in water molecules, we expect much longer length scale for the collective motions of water molecules at THz frequencies. As shown in part by Qi et al. (*58*), molecular dynamics simulations of water confined in carbon nanotubes showed an anomalous decrease in terahertz permittivity at nanotube diameters as large as 7.80 nm. Furthermore, a study in infrared wavelengths has shown that water confined between two metallic surfaces reduced their refractive index by 40 % at gap widths near 10 nm (*45*). Thus, the confinement effect may occur at much larger gap widths at THz frequencies than in the static regime (*65*). Figure 5D illustrates the suppression of long-range dynamics by spatial constraints within metallic nanogaps, which leads to reduced terahertz permittivity of residual water. Figures 5E and 5F indicate a gradual decrease in the complex refractive indices of the residual water at thickness of around 10 nm and below, consistent with Fig. 4. In the smallest gap at THz frequencies, the refractive index decreases by 40%, similar to the trend observed at infrared frequencies (*45*), while the extinction coefficient changes more dramatically and reaches zero. Considering that the imaginary permittivity of water at THz frequencies is dominated by the slow Debye process, which also accounts for the huge static permittivity of water (*57*), the dynamics of nanoconfined water at terahertz frequencies uniquely bridges the two completely different regimes of infrared and microwave frequencies.

**Ice-to-water phase transition under nanoconfinement**

We can draw a parallel between the lowering refractive index observed in nanoconfined water and the changing refractive index during the phase transition from liquid water to solid ice. In the solid state of ice, the spacing between hydrogen bonds becomes fixed, causing water molecules to align in a structured manner. This arrangement results in lower refractive indices compared to liquid water. Therefore, temperature-dependent THz transmission measurements on water-filled nanogaps can provide additional insight on the gap width-dependent collective dynamics of water molecules.

THz time traces and transmitted amplitude spectra of water-filled nanogaps with the gap widths of 5 and 20 nm are shown in Fig. 6A and 6B, respectively. Considering that ice has a smaller complex refractive index than liquid water, the transmission spectra through both samples exhibits a decrease in amplitude and a spectral redshift as the temperature rises from 250 K, surpassing 273 K, to 290 K (see Supplementary Materials Fig. S8). The series of transmitted amplitude changes for each sample is compared in Fig. 6C as a function of temperature. Interestingly, as the confined water phase transitions from ice to liquid, the 5 nm gap exhibits a gradual decrease in transmitted amplitude, whereas the 20 nm gap experiences an abrupt change like bulk water. Despite the 5 nm gap sample being more sensitive to variations in the dielectric environment within, the amplitude change is only half as large as observed in the 20 nm gap. When the residual water is treated as ordinary bulk water, the spectral profile should represent distinctive changes due to the gap-plasmon effect, regardless of the interfacial regions. In Fig. 6D, while the observed spectral changes in the 20 nm gap are quantitatively consistent with calculations using typical refractive index values for liquid water and ice, the changes in the 5 nm gap deviate by approximately three times from the predicted value. Accordingly, the nanoconfined water within the 5 nm gap can be considered somewhat 'ice-like' operating at terahertz frequencies, suggesting that suppressed collective dynamics of water molecules are observed over long distances, as expected. However, it is noted that this does not signify a change in the phase transition temperature of water. For such a shift to occur, confinements in multiple directions and smaller gap widths would be required, as observed in carbon nanotubes (*24*).

**Discussion**

In this study, we performed gap-width dependent terahertz transmission measurements on water-filled metal nanogap to determine refractive indices of the nanoconfined water at THz frequencies. We observed a considerable decrease in both real and imaginary refractive indices of the nanoconfined water compared to those of bulk water, but with a trend that could not be explained sufficiently with the model proposed in the static regime. We attribute this deviation to the suppressed dynamics of the long-range correlation of hydrogen bond networks corresponding to the THz frequency regime under confined volume. This constraint leads to a reduction in the terahertz permittivity of the 'non-interfacial' water, as demonstrated through temperature-dependent measurements. Even though this study focuses on terahertz frequencies, the scope of the study can be expanded to visible, infrared, and even microwave frequencies simply by altering the dimensions of the metal nanogaps. It is also possible to study interfacial dynamics of water molecules on different metals, or even on dielectrics or self-assembled molecules by coating the layers after fully etching the gap. Furthermore, if the water-filled gap can be realized at sub-nanometer gap widths (*56*), our understanding of nanoconfined water may be extended to much more exotic phenomena, such as various phases of monolayer (*19, 66*) and an anomalous shift of phase transition temperatures (*24*). Therefore, our scheme provides a new method to study and utilize water-mediated processes such as protein folding, lipid rafts and molecular recognition.

**Materials and Methods**
**Fabrication of water-filled nanogap loops**

First, we cleaned a 500-μm thick silicon substrate using acetone and isopropyl alcohol (IPA) in an ultrasonic environment to remove any particles. We then applied image reversal photoresist (AZ 5214E) onto the silicon wafer at 4000 rpm for 30 s, followed by a soft bake at 105 °C for 90 s to eliminate residual solvent. The coated sample was exposed to UV light with an intensity of 10 mW/cm² and a wavelength of 365 nm (i-line) for 5 s using a MIDAS MA-6 mask aligner. After the reversal bake and flood exposure, we developed the sample with MIF 300 developer for 120 s and rinsed it with DI water for an additional 120 s. Subsequently, we deposited a 200 nm-thick layer of gold and a 2 nm-thick layer of chromium (Cr) using an electron beam evaporator, and then removed any remaining photoresist with acetone. Employing an atomic layer deposition system (Lucida 100), we then deposited a conformal layer of $Al_2O_3$, varying from 2 to 20 nm, using trimethylaluminum (TMA) as the precursor and DI water as the reactor, controlling the thickness by adjusting the number of cycles, all at a temperature of 200 °C. The deposition rate of $Al_2O_3$ was 1.1 Å per cycle, with the thickness confirmed using ellipsometry. Afterwards, a 200 nm-thick layer of gold was deposited on top of the sample, which filled the empty region and defined the metal-insulator-metal gap. Excess gold that was deposited on top of the first gold layer was subsequently removed using adhesive tape. To achieve a smoother surface, we polished the nanogap loops' exteriors using an Ar-ion miller at an oblique angle of 85 degrees for 5 minutes. Following this, we etched the $Al_2O_3$ inside the nanogaps using a 1 M KOH solution, varying the etching time from 1 minute for the 2 nm gap to 10 minutes for the 20 nm gap, depending on the gap width. Following the etching process, we immediately transferred the KOH bath to a DI water (resistance = 18.2 MΩ· cm) reservoir, allowing DI water to dilute the KOH solution and simultaneously fill the gap. Finally, we sealed the water-filled nanogap loops using a sapphire cover (650 μm) with double-sided Kapton tapes (100 μm) as a spacer within a DI water bath. For low temperature measurements, we additionally sealed sidewalls of the sample using GE varnish, preventing the vaporization of water molecules under ultra-high vacuum conditions.

**Terahertz time-domain spectroscopy**

To investigate the dynamics of water molecules within nanogaps, we employed transmission-type THz time-domain spectroscopy. We utilized a commercial Ti:sapphire oscillator system (Synergy BB, Spectra-Physics) operating at a central wavelength of 780 nm, producing sub-10 fs pulse widths with a repetition rate of 75.1 MHz. For terahertz generation, we employed a commercial GaAs photoconductive antenna (Tera-SED3, Laser Quantum) with a field amplitude of 300 V/cm. The resulting THz pulse spanned a spectral range from 0.1 THz to 3 THz and was guided using a series of off-axis parabolic mirrors. The focused terahertz pulse, with a beam size of 2 mm, was directed onto the water-filled nanogap loops. Detection of the transmitted terahertz pulse was accomplished through an electro-optic sampling method utilizing a 1 mm-thick (110) ZnTe crystal. To mitigate water vapor absorption effects at terahertz frequencies, we maintained a dry air purging system throughout the experiment.

**Temperature-dependent measurement**

We employed a liquid helium (He) flowed cryostat (MicrostatHe2, Oxford Instruments) to confirm the phase transition of water within the nanogap. To avoid multiple reflections of terahertz waves, we utilized a transparent 3 mm-thick TPX window optimized for terahertz frequencies. Furthermore, we designed a customized holder from oxygen-free copper, renowned for its superior thermal conductivity. This holder incorporated a temperature sensor (DT-670, Lakeshore) to ensure precise temperature measurement of the sample. Heat transfer was facilitated through a wire, with a cryogenic manganin wire of exceptionally low thermal conductivity employed to impede heat flow. A segment of this wire was attached to a cold finger for effective heat anchoring. This meticulous arrangement yielded a temperature difference between the cold finger and the sample part within a remarkably tight ±0.1 K range in the low-temperature domain. For affixing the sample to the specific holder, we utilized GE varnish, allowing ample drying time of 24 hours prior to loading it into the cryostat. All low-temperature measurements were carried out under high vacuum conditions ($< 8.5 \times 10^{-7}$ mbar) to prevent the freezing of water vapor.

## Acknowledgments

**Funding**
This work was supported by the National Research Foundation (NRF) of Korea Government (MSIT: NRF-2015R1A3A2031768, NRF-2020R1A2C3013454, NRF-2021R1A2C1008452, NRF-2021R1C1C1010660, NRF-2022M3H4A1A04096465), the Republic of Korea's MSIT (Ministry of Science and ICT) under the ITRC (Information Technology Research Center) support program (IITP-2023-RS-2023-00259676) supervised by the IITP (Institute of Information and Communications Technology Planning & Evaluation), 2023 Research Fund (1.230022.01) of Ulsan National Institute of Science and Technology (UNIST), and Semiconductor R&D Support Project through the Gangwon Technopark(GWTP) funded by Gangwon Province(No. GWTP 2023-027).

**Author contributions:** D.S.K. conceived the original idea. H.Y., G.J. and J.J. fabricated the samples and performed the THz-TDS measurement and analysis. G.J., K.K., and H.P. built the temperature-dependent THz-TDS setup. S.P. and Y.D.P. built the CPD setup, carried out measurements, and analyzed. H.A. and J.W.J. built the CA setup, carried out measurements, and analyzed. M.C. and N.P. performed the QM/MM calculations. H.T.L calculated the bulk indices from the experimental results. All authors analyzed the data and discussed the results. All authors wrote and approved the final draft of the study. H.P., D.S.K., and J.J. supervised the project.

**Competing interests:** The authors declare no conflicts of interest regarding this article.

**Data and materials availability:** All data, code, and materials used in the analyses must be available in some form to any researcher for purposes of reproducing or extending the analyses. Include a note explaining any restrictions on materials, such as materials transfer agreements (MTAs). Include accession numbers to any data relevant to the paper and deposited in a public database; include a brief description of the dataset or model with the number. The DMA statement should include the following: "All data are available in the main text or the supplementary materials."

# Figures

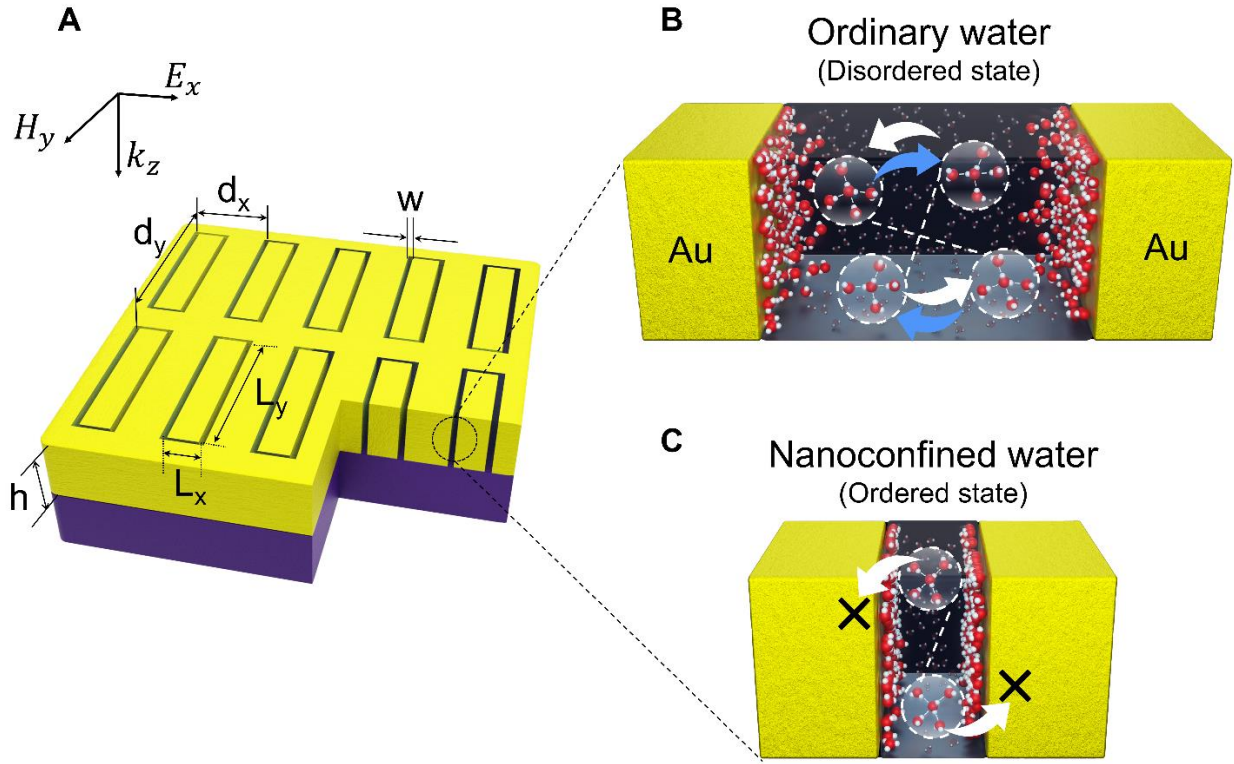

**Fig. 1. Rectangular loop nanogaps filled with water.** (**A**) Schematic diagram of a p-polarized THz electromagnetic wave incident on the confined water in metal nanogap loops for measuring long-range correlation of hydrogen bond networks depending on the gap width. The gap width $w$ varies from 2 to 20 nm in different samples with the same metal thickness of $h = 200$ nm. Parameters of the nanogap loop array are as follows: $L_x$ and $L_y$ are the horizontal and vertical lengths of the rectangular loop, and $d_x$ and $d_y$ are the horizontal and vertical periods of the loop array. (**B**, **C**) The concepts of ordinary water and nanoconfined water are illustrated in (B) and (C), respectively.

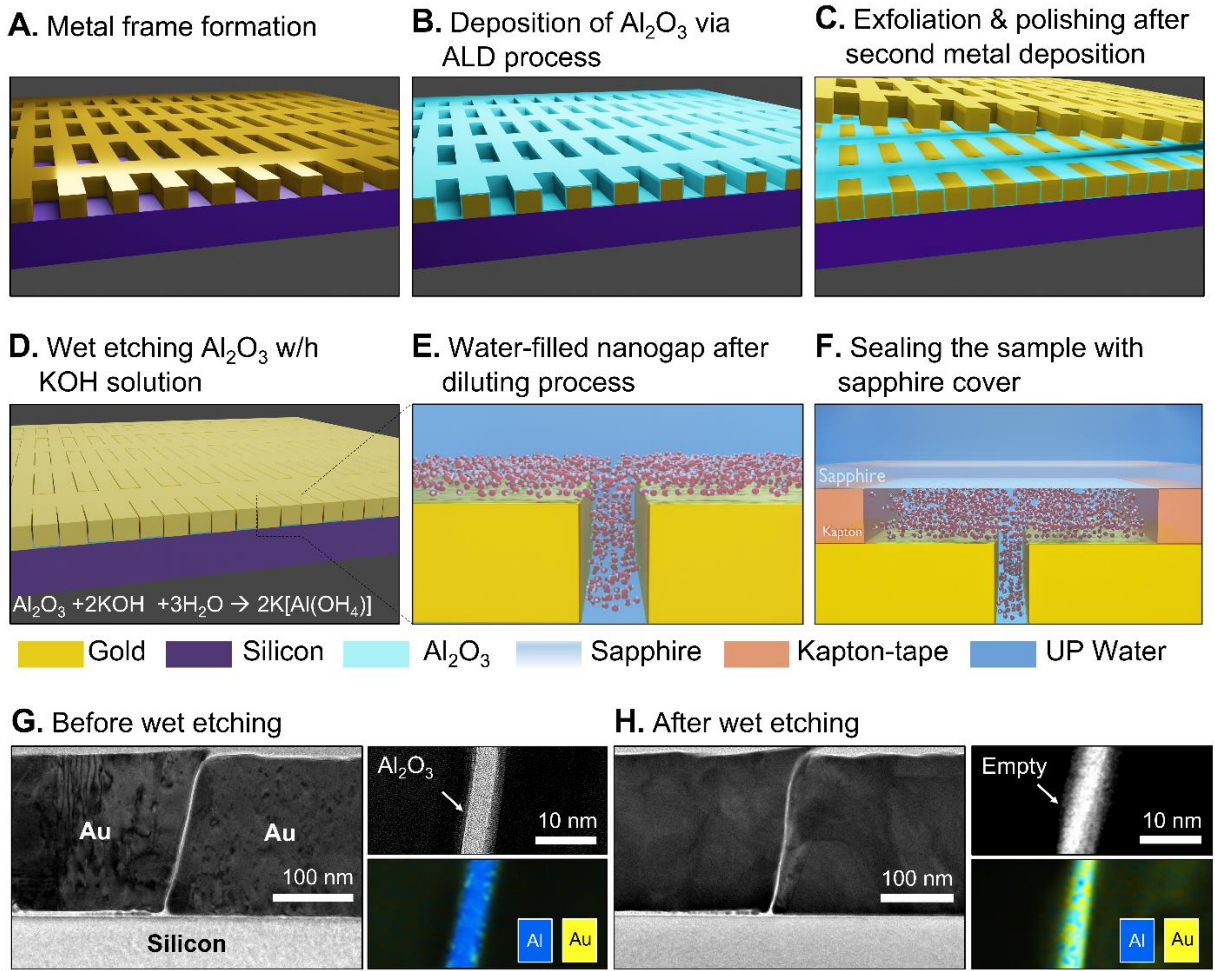

**Fig. 2. Fabrication process of water-filled nanogap loops.** (**A**) A gold frame patterned using conventional photolithography and lift-off process. (**B**) Atomic layer deposition of aluminum oxide at thicknesses ranging from 2 to 20 nm. (**C**) Exfoliation of excess metals following deposition of the second gold layer reveals planarized metal nanogaps with gap widths ranging from 2 to 20 nanometers. (**D**) By wet-etching the gap-filling alumina oxide, the gap is filled with water and reaction intermediates. (**E**) By diluting the gap-filling solution using deionized (DI) water, water-filled nanogaps can be achieved. (**F**) Water-filled nanogaps are sealed with a sapphire cover and a double-sided Kapton tape. (**G, H**) Cross-sectional scanning transmission electron microscopy (TEM) images and energy dispersive spectroscopy (EDS) maps of the 5 nm gap (G) before and (H) after wet etching are shown.

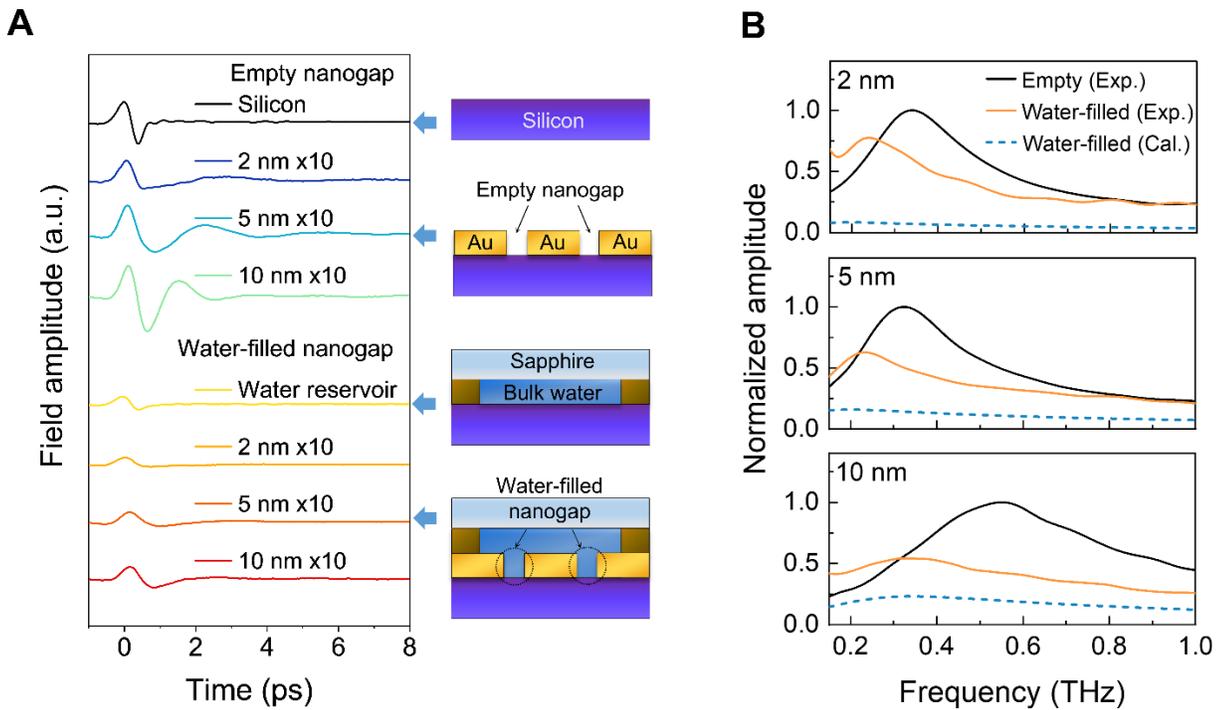

**Fig. 3. THz transmissions of empty and water-filled nanogaps with different gap widths.** (**A**) Time traces of terahertz pulses transmitted through nanogap samples with empty and water-filled gaps, as well as through a bare silicon substrate with and without a water reservoir. The time traces have been vertically offset for clarity. The time traces of 2, 5, and 10 nm gaps have been magnified 10 times to enhance visibility and highlight their small-scale fluctuations. As indicated by the arrows, each sample schematic is shown in the right panel of the graph. (**B**) Normalized THz amplitude (transmitted electric field) spectra of empty (black solid) and water-filled (orange solid) nanogaps with the gap widths of 2, 5, and 10 nm, obtained by the THz time-domain spectroscopy. The blue dashed lines represent the normalized field spectra, calculated analytically by using the coupled-mode method assuming that the gap-filling water is bulk-like. In all cases, nanogap loop arrays contain the same parameters as follows: $L_x = 20$ μm, $L_y = 80$ μm, $d_x = 40$ μm, and $d_y = 100$ μm

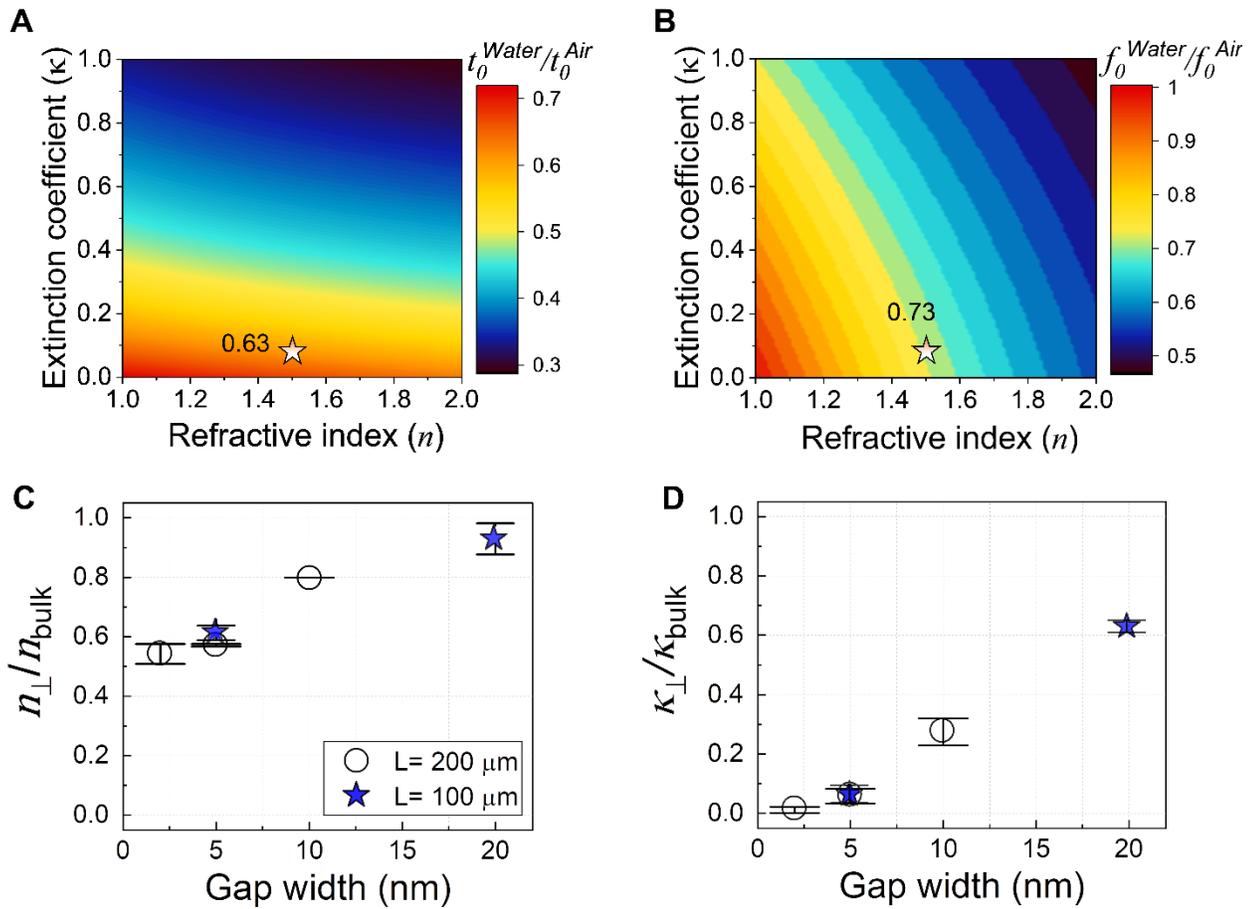

**Fig. 4. Quantitative estimation of the complex refractive index, $\tilde{n} = n + i\kappa$, of gap-filling water within a gap width range of 2 to 20 nm.** **(A, B)** We can uniquely assign the refractive index ($n$) and extinction coefficient ($\kappa$) of the gap-filling medium based on 2D maps of the relative resonance peak amplitude (in A) and the relative resonance frequency (in B) with respect to those of an empty gap (i. e., $n = 1, \kappa = 0$). The star marks in (A) and (B) show the relative resonance peak amplitude and relative resonance frequency of 63% and 73%, respectively, for a water-filled gap with a width ($w$) of 5 nm and a loop length ($L = 2(L_x + L_y)$) of 200 mm. **(C, D)** Estimated relative refractive indices ($n_\perp/n_{bulk}$, C) and relative extinction coefficients ($\kappa_\perp/\kappa_{bulk}$, D) of the nanoconfined water as a function of the gap width. Both $n_\perp$ and $\kappa_\perp$ are $x$-direction refractive indices, being perpendicular to the metal surface, namely the sidewall of the first metal pattern shown in Fig. 2A.

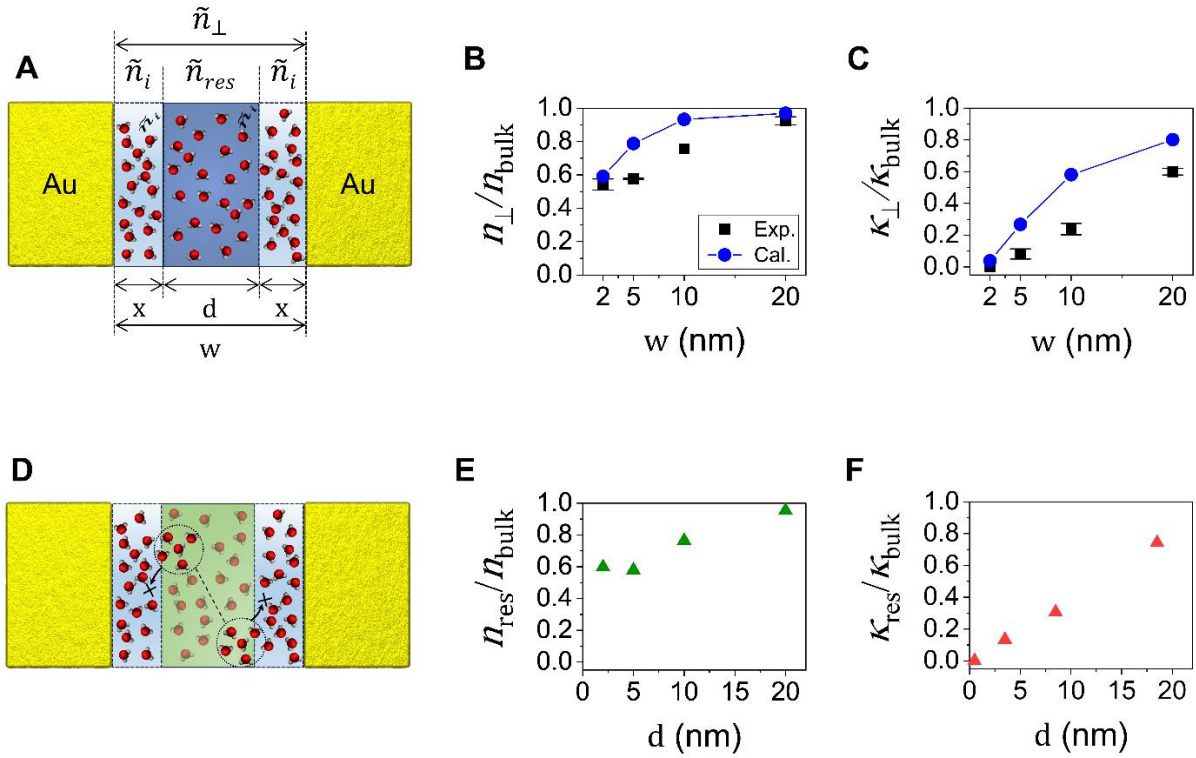

**Fig. 5. Anomalously low refractive index of water confined in metal nanogaps at THz frequencies.** (**A**) A scheme of effective medium theory with different index values for the interfacial and residual water. Symbols denote $\tilde{n}_\perp = n_\perp + i\kappa_\perp$ for water confined in the nanogaps with different gap widths of $w$. It assumes that the gap filling materials are divided in three parts: two interfacial water layers with $\tilde{n}_i = 1.37 + i0$ and a thickness of $x$, and an residual water with $\tilde{n}_{res}(f) = n_{bulk} + i\kappa_{bulk}$ and a thickness of $d$. (**B**, **C**) Blue circles: $n_\perp(w)$ and $\kappa_\perp(w)$ behaviors calculated using the effective medium theory when the residual water has the bulk index. Black squares: Index values extracted from Figs. 4C and 4D are compared with the blue circles. (**D**) A scheme of restricted collective motions of the residual water in metal nanogap. (**E**, **F**) Estimated $n_{res}(w)$ and $\kappa_{res}(w)$ for the residual water with different thicknesses as depicted in the green area in D.

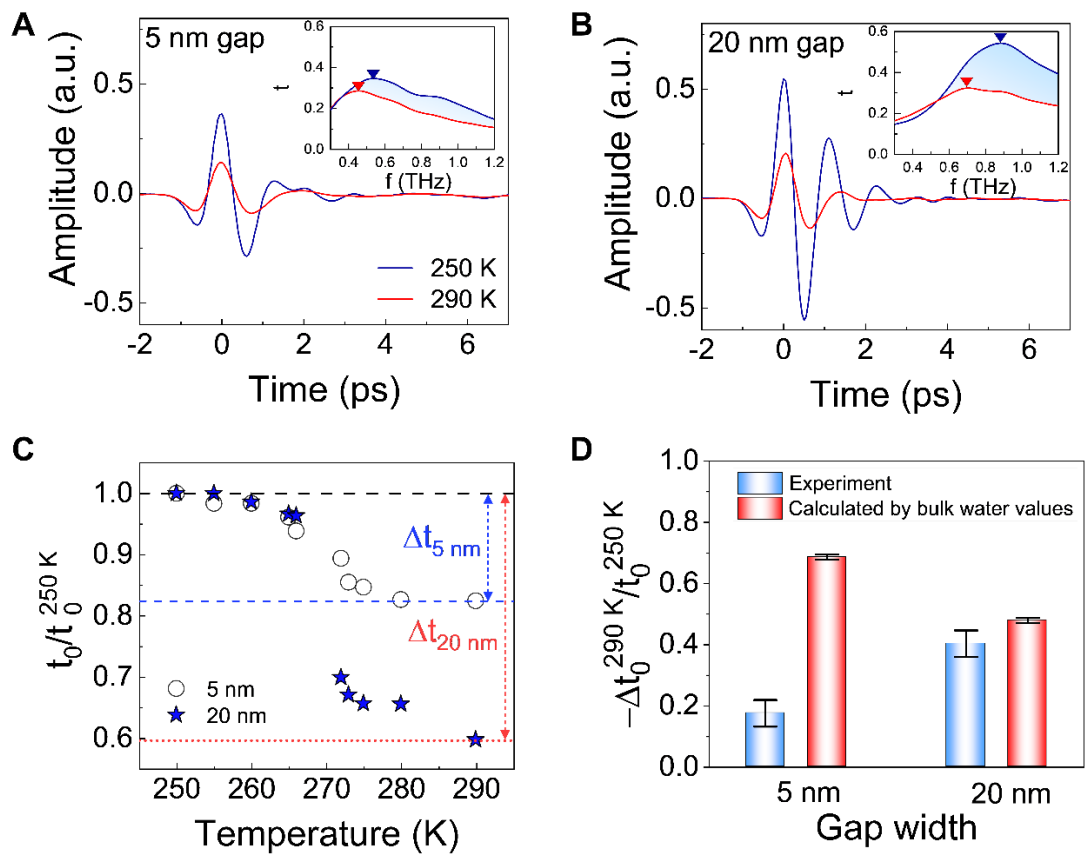

**Fig. 6. Temperature dependent THz time domain spectroscopy with water confined in metal nanogaps.** (**A**, **B**) Time traces of the nanoconfined water from ice to liquid water phase for the gap widths of 5 nm and 20 nm, respectively. Insets in (A, B): Fourier-transformed transmitted amplitude spectra in frequency domain from ice to water for each gap. (**C**) Relative resonance peak amplitudes depending on the temperature for (circle dot) 5 nm and (star dot) 20 nm gap. All the data has been normalized by the peak amplitude at 250 K. (**D**) Comparison of the relative peak amplitude changes between the two phases of liquid water and ice with the gap widths of 5 and 20 nm. (blue bars) Experimental results by temperature dependent THz time domain spectroscopy. (red bars) Calculation results with the bulk water using the coupled-mode method.

# Supplementary Materials for

## Suppressed terahertz dynamics of water confined in nanometer gaps

Hyosim Yang *et al.*

*Corresponding author. Email: Jeeyoon Jeong. peterjjy@kangwon.ac.kr; Dai-Sik Kim. daisikkim@unist.ac.kr; Hyeong-Ryeol Park. nano@unist.ac.kr

**This PDF file includes:**

Supplementary Text
Figs. S1 to S8
Tables S1 to S2
References

**Supplementary Text**

Analytical calculation using coupled-mode method

Numerical simulations including finite-difference-time-domain (FDTD) and finite element method (FEM) are frequently employed to forecast the behavior of electromagnetic waves in sub-wavelength structures in visible and near-infrared wavelengths. However, when simulating nanostructures in terahertz frequencies, there exists a computational limitation due to the disparity in size between the long wavelengths and the sub-wavelength structures. Therefore, for theoretical analysis of metal nanogaps (Fig. S1), we instead perform analytic calculations using the coupled-mode method developed by Garcia-Vidal et al., considering gap plasmon effects (1, 2). For a metal-insulator-metal (MIM) structure, if the gap size is sufficiently small, dispersion relation of the gap plasmon can be expressed as (3),

$$\tanh\frac{k_d w}{2} = -\frac{k_m \varepsilon_d}{k_d \varepsilon_m}, \quad k_i^2 = \beta^2 - k_0^2 \varepsilon_i \, (i=d,m)$$

Here, $k_i$ ($i=d,m$) is wavevector component normal to the metal-dielectric interface where $d$ and $m$ denote dielectric and metal, $w$ is the gap width, $\beta$ is the propagation constant, and $\varepsilon_i$ ($i=d,m$) is dielectric permittivity of the dielectric or the metal. For $\varepsilon_m \gg \varepsilon_d$ the equation reduces to: $k_d^2 / k_0^2 \simeq i\delta / w$, where $\delta$ is the skin depth of metal (~100 nm at 1 THz). It implies that when $w$ is smaller than the skin depth, gap plasmons start to dominate and mostly determine spectral features of the nano-slots such as resonance frequency and transmitted amplitudes.

It should also be noted that while our nanogap is in the shape of a rectangular loop, it can be approximated as two rectangular slots separated by $d_x$. When a terahertz electric field is perpendicular to the long side ($l$) of the rectangular loop, the lowest TE waveguide mode (TE$_{11}$) is excited, and the electric field distribution is sinusoidal along the loop. The electric field along the long axis of the rectangular loop nanogap makes a significant contribution to the far-field transmission, while the short axis exhibits an antisymmetric field distribution, resulting in a zero net contribution to the far-field scattering. Consequently, the electric field distribution in the rectangular loop is nearly identical to that of two separated nanoslots with the same length (4, 5).

As shown in Fig. S1, we define three regions for calculating optical properties of the nanoslot: the (I) below, (II) inside, and (III) above the nanoslot. We assume that terahertz field is incident from region I. Solving the boundary conditions at each interface, we can obtain two coupled linear equations,

$$(G_I - \Sigma)E - G_V E_{gap} = I_0$$

$$(G_{III} - \Sigma)E_{gap} - G_V E = 0$$

where $E$ and $E_{gap}$ are electric fields at the input and output sides of the slots. $E_{gap}$, which is directly proportional to the far field transmission per diffraction theory, can be expressed as,

$$E_{gap} = \frac{I_0 G_V}{(G_I - \Sigma)(G_{III} - \Sigma) - G_V^2},$$

$$I_0 = \frac{i\sqrt{2}}{1+Z_s}\frac{4}{\pi},$$

$$G_{I,III} = \sum_m \sum_n \frac{iwl}{d_x d_y} \frac{\varepsilon_{I,III}(k_0 + Z_s k_{Iz,IIIz}) - k_n^2}{(k_{Iz,IIIz} + Z_s k_0)(k_0 + Z_s k_{Iz,IIIz})} \text{sinc}^2(\frac{wk_m}{2})\left[\text{sinc}(\frac{lk_n + \pi}{2}) + \text{sinc}(\frac{lk_n - \pi}{2})\right]^2,$$

$$G_V = \frac{1}{k_0 h (1 + \frac{k_x^2}{\varepsilon_d k_0^2}) \times \text{sinc}(k_z h)},$$

$$\Sigma = \cos(k_z h) \times G_V$$

where $I_0$ represents the external illumination on the slot under normal incidence condition, $G_{I, III}$ are the coupling constants of the gap mode with regions I and III, $\Sigma$, $G_V$ are constants that depend on the propagation constant of the gap, $Z_s = \sqrt{\varepsilon_m}$ is surface impedance of the metal, $k_m = 2\pi m/d_x$, $k_n = 2\pi n/d_y$, and $k_{Iz,IIIz} = \sqrt{\varepsilon_{I,III} k_0 - k_m^2 - k_n^2}$. Summation over ~100 indices of $m$ and $n$ leads to a reliable, converged result for and takes about one minute to obtain one spectrum.

Complex refractive indices of bulk water in terahertz frequency

To extract the complex refractive indices of bulk water, we measured terahertz transmission through 100 μm-thick water layer sandwiched between a silicon substrate and a sapphire sealing. As multiple reflections of terahertz waves can occur in the water-silicon and water-sapphire interfaces, normalized transmitted amplitude may be expressed as,

$$t = \frac{t_{\text{Water reservoir}}}{t_{\text{Air reservoir}}} = \frac{(n_{Sapphire} + n_{Air})(n_{Air} + n_{Si})n_{Water}}{(n_{Sapphire} + n_{Water})(n_{Water} + n_{Si})n_{Air}} \times e^{i2\pi f (n_{Water} - n_{Air})} \times FP$$

Here, $t$ is the complex transmitted field amplitude of 100 μm thick water as shown in Fig. S2**a** and FP represents the Fabry-Perot term which can be expressed as,

$$FP = \frac{(1 - \frac{(n_{Si} - n_{Air})(n_{Sapphire} - n_{Air})}{(n_{Si} + n_{Air})(n_{Sapphire} + n_{Air})} e^{i4\pi f n_{Air} d/c})}{(1 - \frac{(n_{Si} - n_{Water})(n_{Sapphire} - n_{Water})}{(n_{Si} + n_{Water})(n_{Sapphire} + n_{Water})} e^{i4\pi f n_{Water} d/c})}$$

Here, $n_{Air}$, $n_{Si}$ and $n_{Sapphire}$ represent the refractive indices of air, silicon, and sapphire, where $n_{Air} = 1$, $n_{Si} = 3.4$, $n_{Sapphire} = 3.1$, and reservoir thickness $d = 100$ μm. Using the numerical extraction process (*6*), the complex refractive indices of bulk water can be obtained as shown in Figs. S2c and S2d.

The hydrophilicity of the gold surface

To assess the interaction energy between the gold nanogap and water in our experimental setup, we conducted contact angle measurements using deionized (DI) water. The gold film utilized in our experiment was designed to mimic the surface of the metal nanogap. A droplet of DI water with a volume of 1 μl and a diameter of approximately 1 mm was carefully located onto the gold film. Given the size of the droplet smaller than the capillary length, we can disregard the effects of gravity on the measurements (*7, 8*). The resulting contact angle, determined to be approximately 85 degrees, provides a valuable insight into the wetting behavior of the water droplet on the gold film. According to established conventions, contact angles lower than 90 degrees characterize the surface as hydrophilic (*9*). Leveraging this information, we can extrapolate the interaction energy between the gold surface and water (*10*), a critical parameter employed in the subsequent Density Functional Theory (DFT)-based Molecular Dynamics (MD) simulations.

Radial distribution calculations of hydrophobic and hydrophilic surface

We performed a hybrid quantum mechanics/molecular mechanics (QM/MM) (*11*) calculations using GROMACS (*12*) and CP2K (*13*) to simulate water dynamics between two metal plates (with copper). The PBE-type gradient corrected exchange-correlation potentials (*14*) and tip3p model (*15*) in amber99 force-field (*16*) were employed to describe the copper plates using the CP2K package and water using the GROMACS, respectively. As for the configuration of the water system, tetrahedral water (mp-696735) supported by Material Project (*17-30*) was extended 14 times on the x-axis and 13 times on the y-axis. The copper plate was obtained by extending the orthorhombic unit cell of Cu (111) to 21×34×2. The z-axis of the water system was constructed by 5.4 nm, where the distance between the two copper plates. To generate the hydrophilic copper surface, a structure in which the surface of the copper plate extended by orthorhombic Cu (111) to 7×17×2 was covered with hydroxyl groups was optimized using a QUANTUM ESPRESSO package (*31, 32*), and the supercell was used as a 3×2×1 extended structure. Each configuration was annealed from 0 ns to 1 ns at 0 K to 300 K, equilibrated to 2 ns at 300 K, and simulated at 300 K to 7 ns.

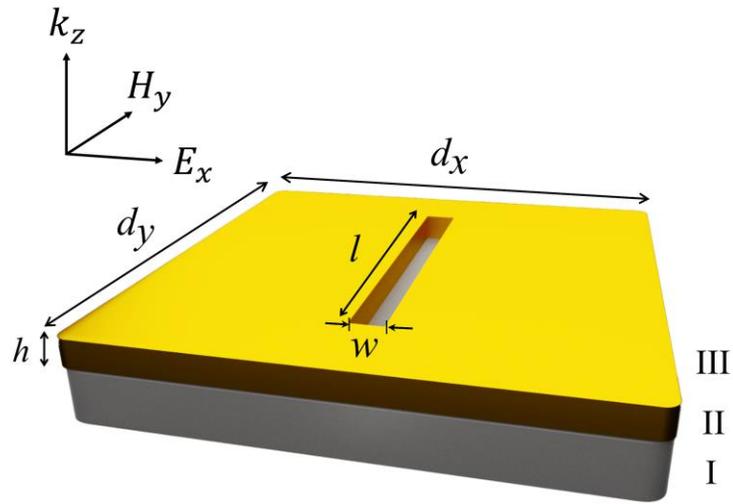

**Fig. S1.** The unit cell of the metal nanogap used in theoretical calculation.

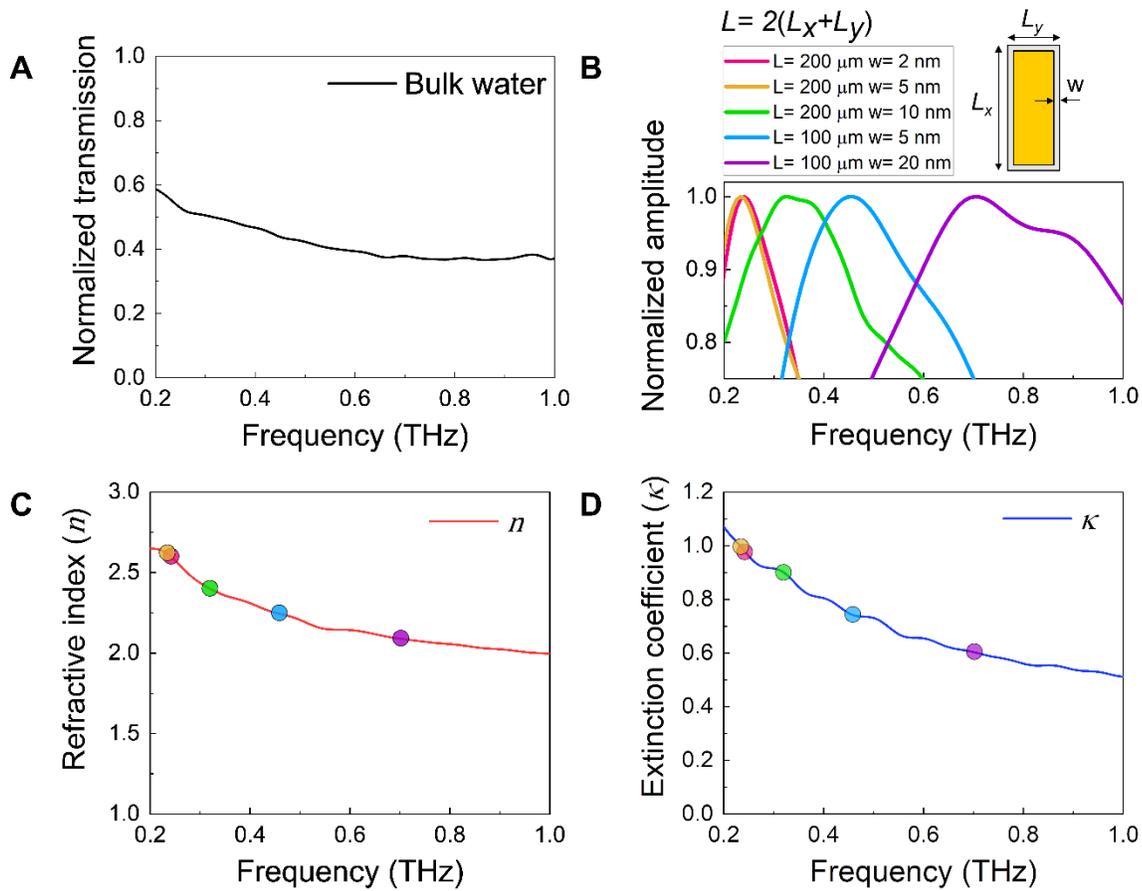

**Fig. S2.** (**A**) Terahertz transmitted amplitude spectrum of a 100 μm-thick bulk water. (**B**) The resonance frequencies of water-filled nanogaps with different loop lengths and gap widths. (**C, D**) Extracted refractive index of $n$ (C) and extinction coefficient of $k$ (D) of 100 μm-thick bulk water in the frequency range between 0.2 and 1.0 THz. In Fig. S2C and S2D, the dots represent the refractive indices and the extinction coefficients corresponding to the resonance frequencies in Fig. S2B, where the color of the dots indicates the geometry of the nanogap loops in Fig. S2B. Since the complex refractive index of bulk water has a dispersion, we used the corresponding index value at the resonance frequency of each nanogap shown in Fig. S2B for coupled-mode calculations.

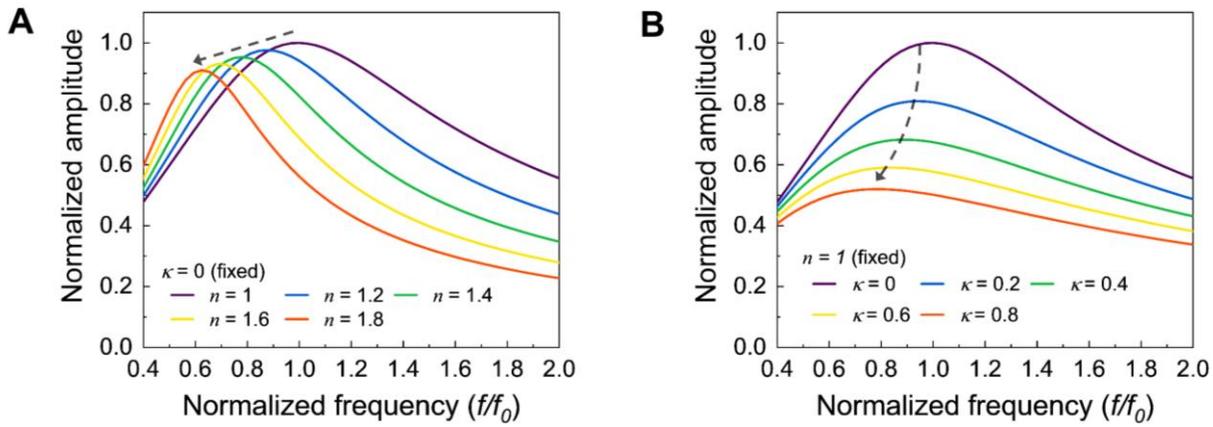

**Fig. S3.** Calculated terahertz transmission amplitude spectra of the nanogap loops with the gap width of 5 nm and the loop length of L = 200 μm. We performed a series of calculations with varying real and imaginary refractive indices of the gap-filling medium. The peak amplitude and resonance frequency of the calculated spectra are normalized with respect to those of an empty nanogap ($n = 1$, $k = 0$). (**A**) With increasing refractive index $n$ with a fixed $k = 0$, the transmission spectrum shows a large redshift of the resonance with a small decrease in amplitude. (**B**) With a fixed $n = 1$, increase of extinction coefficient $k$ shows a large decrease in the transmitted amplitude with a small redshift of the resonance frequency.

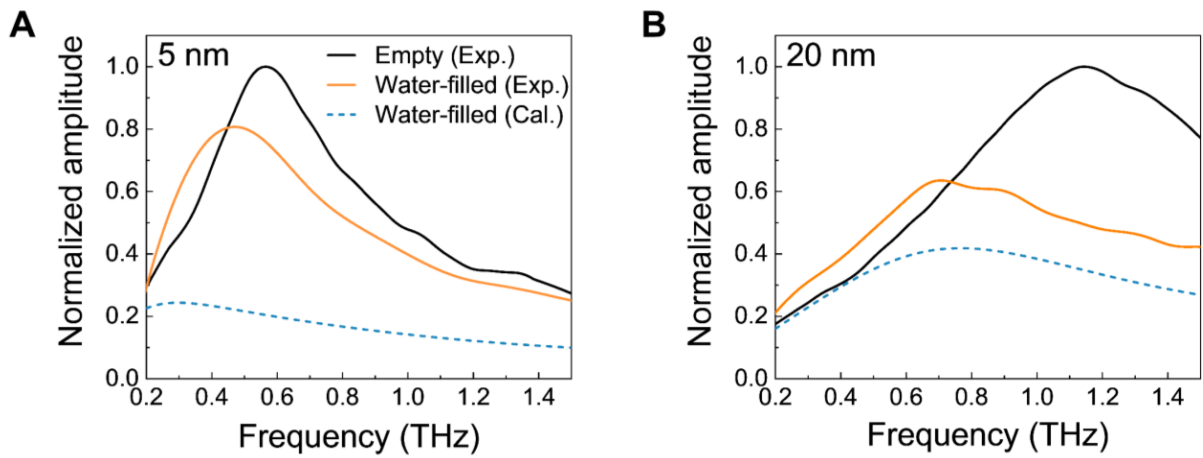

**Fig. S4.** Normalized terahertz transmitted amplitude spectra for empty (black solid) and water-filled (orange solid) metal nanogaps with the gap widths of (**A**) 5 nm and (**B**) 20 nm, and the loop length $L = 100$ μm. The blue dashed line represents the calculated transmission spectra with bulk water value.

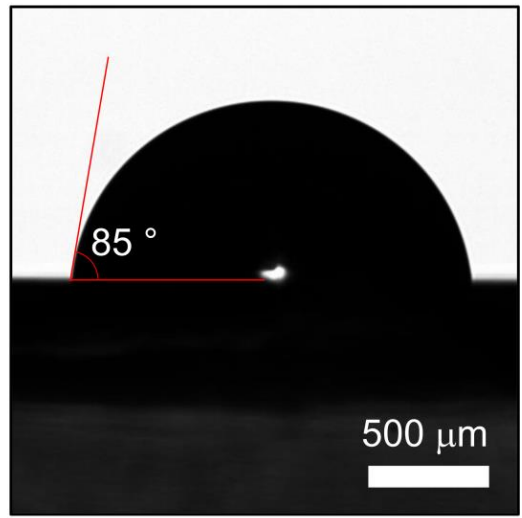

**Fig. S5.** The measured contact angle of a water droplet (1 μl) on a 200 nm-thick gold surface after the wet etching process for the Al$_2$O$_3$ layer.

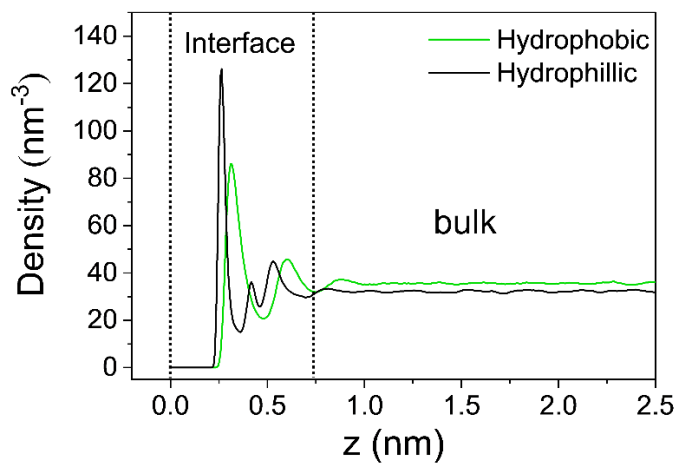

**Fig. S6.** The water density profile of 5 nm gap on hydrophobic surfaces (illustrated in green) and hydrophilic surfaces (illustrated in black) as a function of distance "z" from the surface metal. To achieve a hydrophilic surface, we introduce hydroxyl groups onto the copper wall, resulting in an interfacial water thickness of approximately 0.75 nm. The hydrophobic nature of the copper wall leads to a more extended interfacial water thickness of around ~0.82 nm.

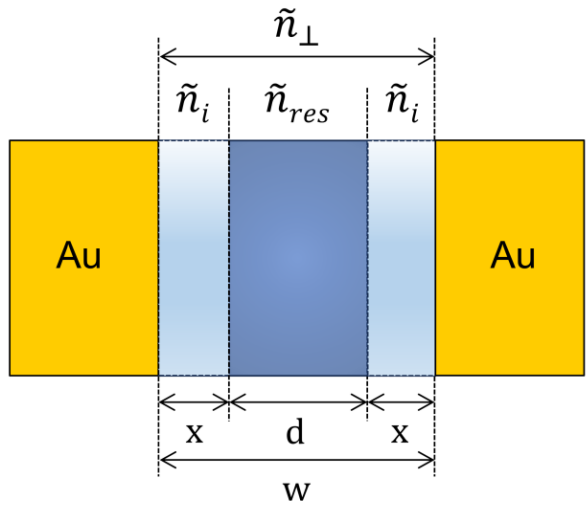

**Fig. S7.** Sketch of our metal nanogap geometry. Here, w represents the width of the nanogap, x is the thickness of interfacial water, and d is the thickness of the residual water. The corresponding THz refractive indices are denoted as $\tilde{n}_\perp$, $\tilde{n}_i$, and $\tilde{n}_{res}$ respectively and $L$ represents the perimeter of the loop nanogap. $n_{bulk}$ and $\kappa_{bulk}$ are refractive indices of bulk water at each resonance of the nanogap loops.

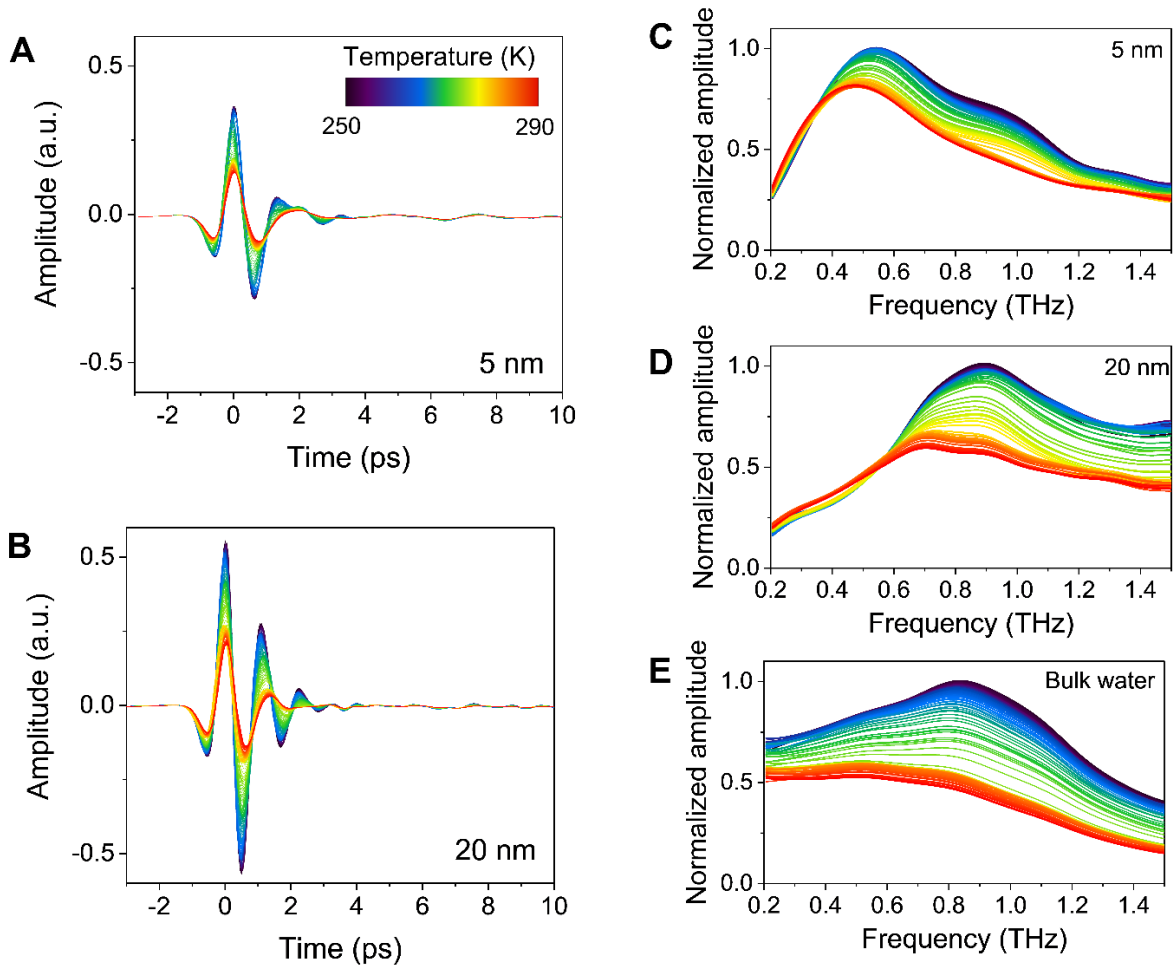

**Fig. S8.** Terahertz time traces of the water-filled nanogaps with the gap widths of (**A**) 5 and (**B**) 20 nm, depending on the temperature from 250 to 290 K. Corresponding normalized terahertz transmitted amplitude spectra of the water-filled nanogaps (**C** 5 and **D** 20 nm gaps) and e the bulk water. Temperature scale bar of all measurement data (THz time traces and normalized amplitude spectra) is displayed in Fig. S8**A** inset.

**Table S1.** Comparison of the real parts of refractive index for the gap-filling, residual, and bulk water at different resonances of the nanogap loops.

| | | $n_i = 1.37$, x = 0.75 nm | | |
|---|---|---|---|---|
| w (nm) | L (μm) | $n_\perp$ | $n_{res}$ | $n_{bulk}$ |
| 2 | 200 | 1.41 | 1.55 | 2.60 |
| 5 | 200 | 1.50 | 1.51 | 2.61 |
| 10 | 200 | 1.74 | 1.83 | 2.40 |
| 20 | 100 | 1.94 | 1.98 | 2.08 |

**Table S2.** Comparison of the imaginary parts of refractive index for the gap-filling, residual, and bulk water at different resonances of the nanogap loops.

| | | $\kappa_i = 0$, x = 0.75 nm | | |
|---|---|---|---|---|
| w (nm) | L (μm) | $\kappa_\perp$ | $\kappa_{res}$ | $\kappa_{bulk}$ |
| 2 | 200 | ~0 | ~0 | 0.98 |
| 5 | 200 | 0.082 | 0.13 | 1 |
| 10 | 200 | 0.215 | 0.27 | 0.89 |
| 20 | 100 | 0.36 | 0.44 | 0.6 |